\documentclass[aps, prl, manuscript, amsmath, showkeys, showpacs]{revtex4-1}

\usepackage{graphicx}

\begin{document}

\title{A mechanism for magnetic field stochastization and energy release during an edge pedestal collapse}

\author{T. Rhee}
\affiliation{National Fusion Research Institute, Daejeon 305-333, 
Republic of Korea}
\author{S. S. Kim}
\affiliation{National Fusion Research Institute, Daejeon 305-333, 
Republic of Korea}
\author{Hogun Jhang}
\email[Corresponding author. ]{hgjhang@nfri.re.kr}
\affiliation{National Fusion Research Institute, Daejeon 305-333, 
Republic of Korea}
\author{G. Y. Park}
\affiliation{National Fusion Research Institute, Daejeon 305-333, 
Republic of Korea}
\author{R. Singh}
\affiliation{National Fusion Research Institute, Daejeon 305-333, 
Republic of Korea}
%
%\date{\today}

\begin{abstract}
On the basis of 
three-dimensional nonlinear magnetohydrodynamic simulations, 
we propose a new dynamical process leading to the 
stochastization of magnetic fields during an edge pedestal collapse.
Primary tearing modes are shown to grow by 
extracting kinetic energy of unstable ballooning modes, 
eventually leading to the island overlap.
Secondary tearing modes, which are generated
through a coherent nonlinear interaction between 
adjacent ballooning modes, 
play a key role in this process, mediating the energy transfer
between primary ballooning and tearing modes.
Explicit calculations of the parallel energy loss through the stochastic
field lines show that it can be a likely dominant energy loss mechanism 
during an edge pedestal collapse.
\end{abstract}

\pacs{05.45.-a, 52.25.Gj, 52.35.Mw, 52.35.Py}
\keywords{Nonlinear, Tearing modes, Stochastic fields, Collisionless diffusion}
\maketitle

\section{}

%%%%%%%%%%%%%%%%%%%General Introduction%%%%%%%%%%%%%%%%%%
An edge localized mode (ELM) is an instability occurring in the edge pedestal
region of magnetic fusion plasmas. It is, in some sense, an unavoidable
consequence of operating a plasma in an enhanced edge confinement mode, 
{\it i.e.} the H-mode \cite{Wagner}, 
characterized by a steep pressure gradient. 
Large ELMs are accompanied with unacceptably high heat
flux to divertor and plasma facing materials in fusion devices. 
Therefore, elucidating physics mechanisms responsible for the ELM 
crash and ensuing energy losses 
has been a central issue for decades in contemporary plasma physics,
as an effort to avoid or mitigate ELMs.

The present idea on the origin of ELMs is based on 
the destabilization of ideal peeling-ballooning modes \cite{PB}.
Formation and ejection of filamentary structures 
have been observed in both nonlinear MHD simulations \cite{JOREK,BOUT,M3D}
and experiments \cite{MAST,AUG,JET,DIIID,KSTAR} during ELM crashes.
Thus, ELMs have often been associated with these filamentary structures.
A theory has been proposed
on the origin of the filamentary structure based on nonlinear 
evolution of the ballooning mode \cite{Explosive}. However, a recent study
shows that the energy loss due to the filaments is estimated
to be $\lesssim 30 \%$ of the total energy loss \cite{Kirk}. 
Then, the identification of the physical mechanism accounting for 
$\gtrsim 70 \%$ of the ELM energy loss remains as a question.

Nonlinear MHD and gyrofluid simulations have often shown the generation
of stochastic magnetic fields during a simulated edge pedestal collapse,
though the degree of stochastization varies in models being applied
\cite{JOREK,M3D,BS,Xu-PRL}.
Since the triggering instabilities of ELMs are ballooning modes 
possessing the twisting parity in nature, 
it is a puzzle how these stochastic field lines are
generated out of the initial ballooning modes. 
%Note that the direct conversion of 
%ballooning parity to tearing parity is known to be difficult. 
Some nonlinear mechanisms have been invoked to explain
the generation of stochastic fields. However, no detailed
analysis shedding light on the process of 
stochastization is yet available.
%there has been
%no discovery of which nonlinear processes are involved in the stochastization.
In this Letter, we elucidate, on the basis of three-dimensional 
nonlinear reduced MHD simulations, a mechanism leading
to the magnetic field stochastization during an edge pedestal collapse.
Development of a series of nonlinearly driven
tearing modes and ensuing island overlap are shown to be responsible for
the generation of stochastic magnetic fields. In particular, 
we highlight the role of the secondary
tearing mode, which is generated via a coherent nonlinear interaction
between adjacent ballooning modes, 
as an {\it agent} for the energy transfer from 
the unstable ballooning mode to the primary tearing mode. We also show that the 
collisionless parallel energy transport through the stochastic
field lines can be a main energy loss mechanism during the crash.

%%%%%%%%%%%%%%%%%%%%%%%%%%%%%%%%%%%%%%%%%%%%%
%%%%%%%%%%%%%%%%%%%%%%%%%%%%%%%%%%%%%%%%%%%%%

%{Simulation model:} 

We perform edge pedestal collapse simulations
using a three-field reduced MHD model, which consists of evolution equations
for vorticity and pressure, and Ohm's law. 
The computational model is basically the same as that of 
previous studies \cite{Xu-PRL,Xu-NF,Xi}, 
except for the pressure evolution equation, as will be described later.
All simulations are performed by using the BOUT++ framework \cite{BPP}
without sources and sinks.
Basic parameters in the simulation are as follows:
$R_0=3.5~(m)$ is the major radius, $V_A=9.5\times 10^{6}$ $(m/sec)$
the Alfven velocity, 
$S=\mu_0 R_0 V_A/\eta=10^{9}$ the Lundquist number, and 
$S_H=\mu_0 R_0^3 v_A/\eta_H=10^{12}$ the hyper-Lundquist number.
%$\eta=4.18\times
%10^{-8}$ the plasma resistivity, resulting in the Lundquist number,
%$S=\mu_0 R_0 V_A/\eta=10^{8}$.
The computational domain is $-0.6\leq \psi_N \leq 0.2$,
where $\psi_N$ is the normalized poloidal flux with
$\psi_N=0$ corresponding to 
the location of the approximate last closed flux surface. 
%Same boundary conditions as a previous study\cite{Xi} are used.
A monotonic $q$-profile in the range, $1.19 \leq q \leq 4.87$ with
$0.66 \leq s=(r/q)(dq/dr)\leq 6.26$ is used. 
$q$ and $s$ values at the maximum pressure gradient location
($r_{max}$) are 2.1 and 4.35, respectively.
%%%%%%
%%%%%%%
The normalized pressure profile $\alpha=-2\mu_0 q^2 R_0 (dP_0/dr)/B^2=3.86$ 
at $r=r_{max}$ while the critical $\alpha$ is $\alpha_c=2.75$.
Thus, the initial pressure profile is unstable
to the peeling-ballooning mode.
Simulations can be performed using different initial perturbations.
A recent study emphasizes the role of initially unstable multi-modes
%multiply unstable perturbations
%addresses the effects of multiple unstable perturbations
in pedestal collapse \cite{Xi}. In this Letter, we present results
initiated from an unstable single mode
because it clearly reveals the nonlinear physics process leading 
to magnetic reconnection. 
We have found that multi-mode simulations yield almost
identical results and conclusions 
presented in this Letter, as long as the initial energy of the unstable 
ballooning mode is strong enough.

Figure~1(a) 
%and (b) 
shows the field line trace at $t=50~\tau_A$ ($\tau_A$: Alfven time), which
typifies a stage just prior to stochastization, 
in the toroidal ($Z$, normalized to 1) and radial ($X=\psi_N$) plane.
Hereafter, we use $\tau_A$ as a unit of time throughout.
An interesting observation can be made in Fig.~1(b) where
field lines corresponding to $n=2n_0$ ($n_0$: toroidal mode number
of the most unstable ballooning mode)
%the second harmonics in the toroidal mode
%number of the most unstable ballooning modes 
are shown at $t=55$.
One can see the growth of magnetic islands,
in the middle of two flux surfaces (dotted line)
where the ballooning modes are most unstable (two solid lines). 
We designate these as the {\it secondary tearing modes} (STM).
Figure~1(c) shows the spatio-temporal evolution of 
the Chirikov parameter ($C$) during a simulation. 
The contour for $C=1$ (the thick black line) indicates that the 
stochastization initiates from $t \approx 60$ 
at just inside the separatrix
where the ballooning mode is most unstable,
propagates almost linearly into the core, and saturates
({\it i.e.,} ceases to propagate) at $t \approx 75$. 
%In the later stage of the collapse, 
Stochastization of magnetic fields after $t \gtrsim 60$ is shown to be
mainly due to the growth of tearing modes with the {\it same} mode number as 
the initial ballooning modes.
We designate these tearing modes and initial ballooning modes
as {\it primary tearing modes} (PTM) and 
{\it primary ballooning modes} (PBM), respectively.

%%%%%%%%%%%%%%%%%%%%%%%%%%%%%%%%%%%%%%%%%%%%%%%%%%%%%%%%%%%%%%%%%%%%%%%
%%%%%%%%%%%%%%%%%%%%%%%%%%%%%%%%%%%%%%%%%%%%%%%%%%%%%%%%%%%%%%%%%%%%%%%
% Generation of tearing partity out of ballooning partiy
%%%%%%%%%%%%%%%%%%%%%%%%%%%%%%%%%%%%%%%%%%%%%%%%%%%%%%%%%%%%%%%%%%%%%%%%
A question then arises as to how 
the even parity tearing modes (STMs and PTMs)
are generated from the odd parity PBMs. 
The ballooning instabilities are known to be difficult to generate magnetic 
reconnection due to their nonlinear self-acceleration property \cite{Cowley}. 
 \begin{figure}
\includegraphics[width=1.0\textwidth]{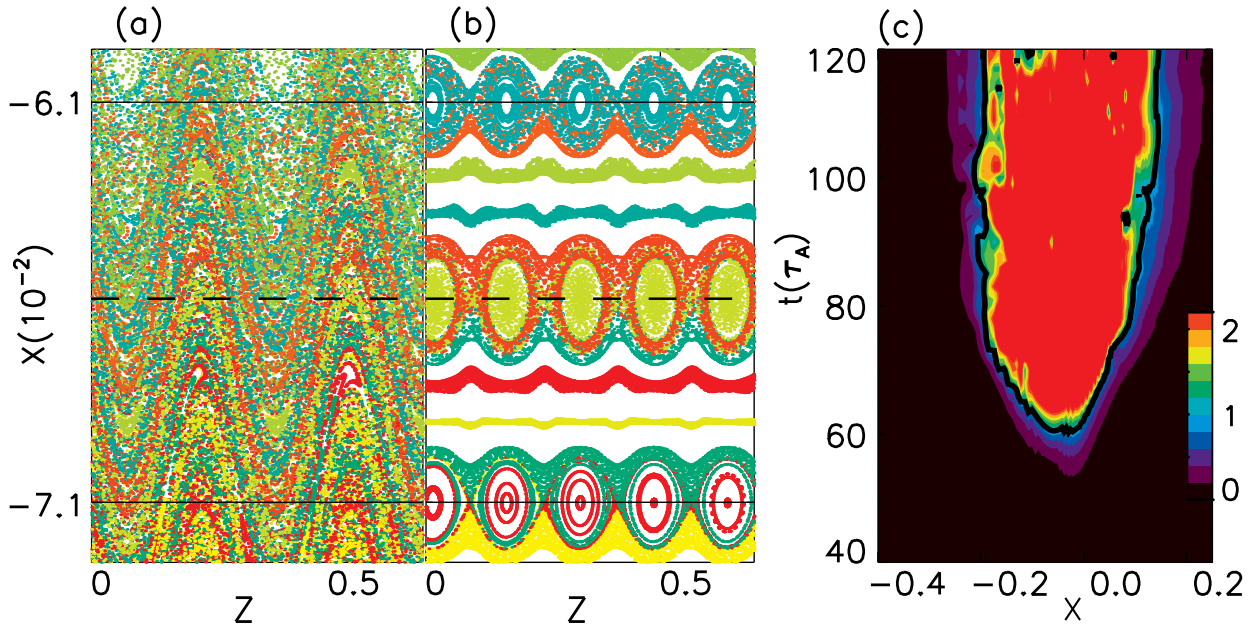} 
\caption{\label{Fig:1} Field line traces at $t=50~\tau_A$ 
($\tau_A$: Alfven time) (a) and $t=55~\tau_A$ 
corresponding to $2n_0$ ($n_0$: the toroidal mode number of
the most unstable ballooning mode) (b) in the 
toroidal ($Z$, normalized to 1) and radial ($X=\psi_N$) plane.
The dotted line represents the midpoint between two flux
surfaces where the ballooning modes are most unstable
(two solid lines).
(c) Spatio-temporal evolution of the Chirikov parameter.}
 \end{figure}
The generation of even parity modes can be seen more clearly in Fig.~2(a) 
where the volume-integrated intensity for the total even 
%($\tilde{A}_{\parallel}^+$, red) 
(${\psi}^+$, red) and odd 
%($\tilde{A}_{\parallel}^-$, black)
(${\psi}^-$, blue)  
components of the perturbed flux (${\psi}$),
$\int_V |{\psi}^{\pm}|^2 dV$, 
are plotted as a function of time. 
Even parity modes start to grow rapidly around 
$t \sim 60$, saturates for $70 \lesssim t \lesssim 85$, then
reduces to a steady state value at $t \sim 105$. 
At final stage, $\int_V |{\psi}^{+}|^2 dV\simeq
\int_V |{\psi}^{-}|^2 dV$, implying the equipartition of
magnetic energy between even and odd parity modes. 
\begin{figure}
\includegraphics[width=1.0\textwidth]{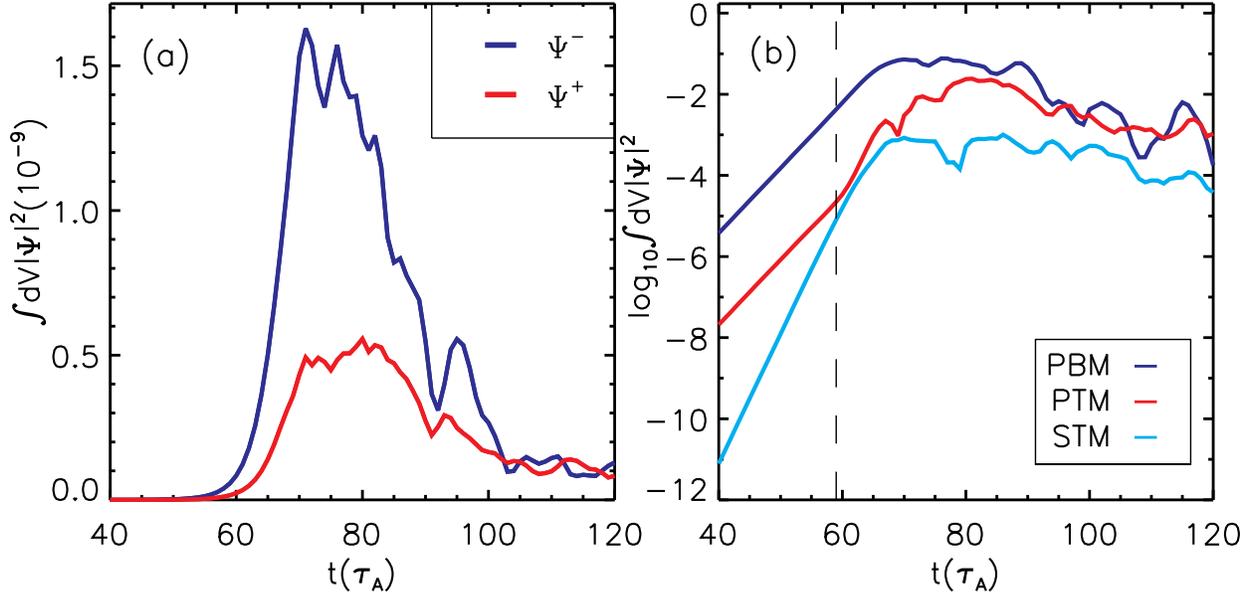} 
\caption{\label{Fig:2} Time evolution of  volume-integrated
intensity for (a)odd (blue) and
even (red) parity modes and (b) PBM (blue), PTM (red), 
and STM (cyan).}
\end{figure}
%%%

Figure~2(b) shows the time evolution of volume-integrated amplitudes for 
the most unstable PBM (blue), PTM (red) and the corresponding STM (cyan). 
In a linear stage ($t \lesssim 60$), both PBM and PTM co-exist and
grow with the same growth rate, even though the magnitude
of PBM is dominant. The initial even parity mode is
parasitic and irrelevant to the actual tearing mode growth.
%hence no island formation. 
The growth rate of STM is exactly twice that of primary modes, as will
be shown shortly.
We emphasize that this STM is a nonlinearly driven mode which grows
even when
%with
%negative $\Delta^\prime$, {\it i.e.,}
%when  $\Delta^\prime <0$, , {\it i.e.,} 
$\Delta_{STM}^\prime \equiv \left[ (1/\psi_{STM})(d\psi_{STM}/dr)\right]_{-W_s/2}^{W_s/2} < 0$,
where $W_s$ is the island width of the STM.
When the magnitude of STM 
becomes comparable to that of PTM at $t \simeq 60$, the 
growth of PTM is accelerated
This acceleration is shown to be possible only mediated by STM. 
Finally, PTMs give rise to an island overlap and subsequent field
line stochastization.
These observations suggest that magnetic reconnection and
ensuing field line stochastization involves a strong nonlinear
interaction among PBM, PTM and STM.

To study the nonlinear interaction quantitatively, we 
first evaluate time evolution of growth rates  
%of  volume-integrated fluctuations 
for PBM ($\gamma^{-}_0$), PTM ($\gamma^{+}_0)$ and STM ($\gamma^{+}_2)$,
the results of which are shown in Fig.~3(a).
%The growth rate of $\psi^{+}_2$ 
One can see that $\gamma_2^+=2\gamma_0^-$,
%is exactly twice that of PBM, 
implying the STM is driven by a coherent nonlinear interaction between 
adjacent PBMs.
%{\it i.e.,} through 
The beating of adjacent PBMs with the mode numbers
($m_0$, $n_0$) and ($m_0+1$, $n_0$) generates
a STM with the mode number ($2m_0+1$, $2n_0$).
Also, one can observe that a sudden increase of $\gamma_0^+$ around
$t \approx 60$ is synchronized to the drop of $\gamma_2^+$.
This signifies the weakening of STM growth due to
the growth of PTM. 
%
%
%growth of PTM by extracting energy from STM.
%
Figure~3(b) shows time evolution of the island width corresponding
to STM (blue, $W_s$) and PTM (red, $W_p$). 
%Initially, 
$W_s$ grows first and approaches to the inter-surface distance,
$W_s \approx 1/n_0 q^\prime$, at $t \simeq 60$. Then, 
$W_p$ begins to grow rapidly. 
The initial stochastization of magnetic 
fields at this time are due to the island overlap by the STMs, while 
the full stochastization is realized by the PTMs after the initial
stochastization. Thus, we arrive at a provisional conclusion 
that the nonlinear energy transfer among relevant modes
are ultimately responsible for the generation of nonlinearly driven 
tearing modes and leads to the field line stochastization. 

%%%
 \begin{figure}
\includegraphics[width=1.0\textwidth]{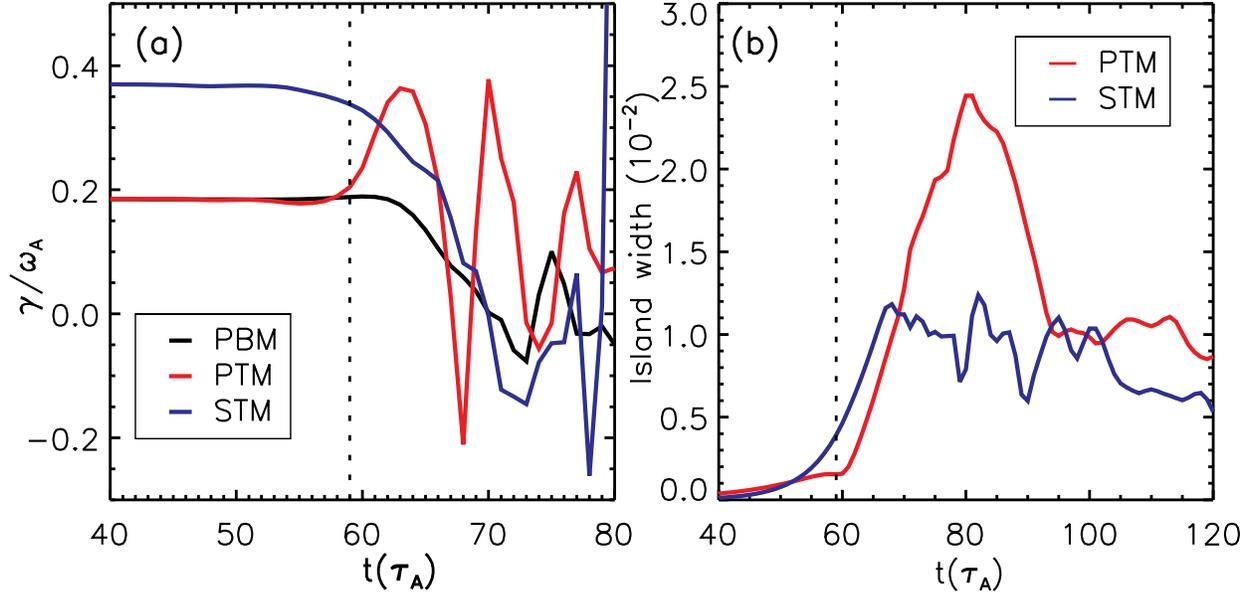} 
\caption{\label{Fig:3} Time evolution of (a) growth rates
for PBM, PTM and STM, and (b) island width corresponding
to STM (blue) and PTM (red).}
\end{figure}
%%%5

For a detailed study of this nonlinear energy transfer process, we 
evaluate the volume integrated nonlinear energy transfer rate
to the magnetic energy of the $(m_0, n_0)$ mode, 
\begin{eqnarray}
\Gamma_{m_0,n_0} = 2{\rm Re}\sum_{m,n} \int_V
J_{m_0,n_0}^+ \left[\phi_{m,n}, \psi_{m_0+m,n_0+n}^* \right]dV,
\end{eqnarray}

\noindent where $J_{m_0,n_0}^+ =  \nabla_\perp^2 \psi_{m_0,n_0}^+$,
and $[a,b]$ denotes the conventional Poisson bracket operation.
$\Gamma_{m_0,n_0}$ represents how much energy is transferred to 
the $(m_0,n_0)$ even parity mode from the kinetic 
%and magnetic energy of 
energy of $(m,n)$ 
%and $(m_0+m,n_0+n$) 
mode represented by $\phi_{m,n}$.
%and $\psi_{m_0+m,n_0+n}^*$, respectively.
%For instance, two adjacent PBMs produce the STM, while 
%Thus, one can recognize that 
Thus, the PTM with the mode number $(m_0,n_0)$ is predominantly generated 
through the coupling between the adjacent PBM and the STM represented by
$\phi_{m_0+1,n_0}$ and $\psi_{2m_0+1,2n_0}^*$, respectively. 

Figure~4(a) shows time evolution of
$\Gamma_{m_0,n_0}$ for STM (red solid) and PTM (black solid).
$\Gamma_{STM}$ starts to rapidly increase around $t \simeq 60$
and reaches to its maximum value at $t \simeq 70$. Then,
$\Gamma_{PTM}$ starts to increase rapidly. 
%by extracting energy from
%STM and PBM at the adjacent surface. 
The dotted line in Fig.~4(a) shows evolution for the
normalized half-width of the magnetic island due to the STM, 
%
%Chirikov parameter corresponding to the STM,
%the Chirikov parameter for STM defined by
%
\begin{eqnarray*}
C_{STM} =\frac{W_{2m_0+1,2n_0}}
%\frac{\left[W_{2m_0+1,2n_0}+W_{2m_0,2n_0}\right]/2}
{2(\psi_{2m_0+1,2n_0}-\psi_{2m_0,2n_0})}.
%\simeq 1,
\end{eqnarray*}

\noindent 
%where $W_{m,n}$ is the island width for
%the mode number $(m,n)$.
%$C_{STM}=1$ 
A strong energy transfer to PTM, which is represented by a sudden
increase of $\Gamma_{m_0,n_0}$ at $t \simeq 70$, occurs when
the island due to STM approaches to the $(m_0,n_0)$ flux surface,
{\it i.e.,} when $C_{STM} \simeq 1$.
Thus, PTM grows by extracting kinetic energy of 
an adjacent PBM {\it via} STM. 
We emphasize here the crucial role of STM as a {\it mediator} 
transferring the PBM kinetic energy to the magnetic energy of PTM, 
as depicted in Fig.~4(b) schematically.
Without excitation of STM, PBM1 cannot directly deliver its energy
to PTM at the location of PBM2. 
We note that this nonlinear energy transfer mechanism
is, in some sense, akin to that of the vortex mode which was studied in 
Refs.~\cite{Waddel} and \cite{Carrerras} where 
nonlinear tearing mode generation is considered from
a single {primary mode}. 
A major difference between the present work and 
Refs.~\cite{Waddel} and \cite{Carrerras} is that
a STM is generated by two PBMs, not by one primary tearing mode 
as in the latter cases, and plays as an
{\it agent} in the nonlinear energy transfer process delivering
the kinetic energy of a PBM to the magnetic energy of 
an adjacent PTM.

%%%
 \begin{figure}
\includegraphics[width=1.0\textwidth]{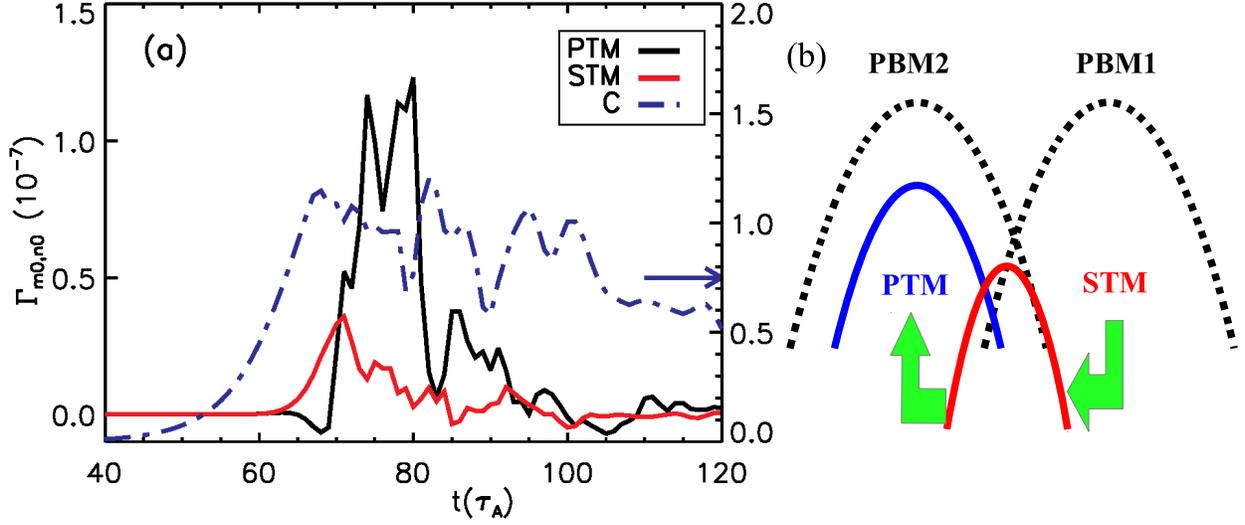} 
\caption{\label{Fig:4} (a) Time evolution of the volume integrated 
nonlinear energy transfer rate ($\Gamma_{m_0, n_0}$) to the magnetic 
energy of STM (red)
%($2m_0+1, 2n_0$, red) 
and PTM (black). %($m_0, n_0$, black). 
$\Gamma_{m_0, n_0}$ is defined in Eq.~(1). 
The dotted line represents the normalized half-width of
the magnetic island due to the STM.
(b) A schematic diagram illustrating the energy transfer
process from the kinetic energy of PBM1 to
the magnetic energy of PTM at the location
of PBM2 {\it via} STM.}  
 \end{figure}
%%%5

Having identified the process leading to magnetic field
stochastization, we now consider collapse-induced energy losses when stochastic magnetic fields are present. 
To this end, we add a term representing heat 
conduction along the stochastic field lines ($Q_{||}$) in the pressure
evolution equation,
\begin{equation}
\frac{\partial P}{\partial t} + \left[ \phi, P \right] =  Q_{||}.
%\mathcal{L}_{||},
%\vec{v}_E \cdot \nabla P =  \mathcal{L}_{||},
\end{equation} 
To find an appropriate model for 
$Q_{||}$, we first evaluate the ratio of Lyapunov length ($L_c$) of 
a magnetic field to the collisional mean free path ($l_{mfp}$) using
the parameters in the simulation, giving rise to
$\lambda \equiv {L_c}/{l_{mfp}} 
%\approx {42 (m)}/{23,000 (m)} 
\simeq 0.01,$
where $L_c$ is evaluated numerically at $t = 70$. 
The whole stochastization process takes place approximately within
$\sim 30~\tau_A$ corresponding to $\sim 0.07$ $\tau_{ei}$ ($\tau_{ei}$:
electron-ion collision time) in this simulation.
The above two conditions indicate that
the stochastization process and ensuing energy losses should be
dealt as a collisionless process. 

An appropriate collisionless fluid model for 
$Q_{||}$ in the presence of stochastic field lines 
is not available at present.
From kinetic simulations, Park {\it et. al.} showed that
the parallel electron heat conduction obeys the 
Rechester-Rosenbluth diffusion model \cite{RR} when one includes a
factor accounting for kinetic effects \cite{GYP}.
Motivated by Ref.~\cite{GYP}, 
we employ the collisionless Rechester-Rosenbluth model to evaluate $Q_{||}$
\begin{equation}
%Q_{||} &=& f_K v_{e}^2 D_{RR} \nabla_\perp^2 P_{00} \nonumber \\
Q_{||} = f_K v_{e}^2 D_{RR} \frac{\partial^2 <P>}{\partial r^2}, %\nonumber \\
%&=& f_K \pi v_{e}^3 R \sum_{m,n}\left(\frac{\delta B_{mn}}{B_T} \right)^2
%\delta_{n,m/q} \frac{\partial^2 P_{00}}{\partial r^2},  
\end{equation} 
\noindent where 
$<P>$ is an equilibrium component
of pressure and $D_{RR}=\pi v_{e} R \sum_{m,n}\left(\delta B_{mn}/{B_T} \right)^2 \delta_{n,m/q}$ with
$v_{e}$ the electron thermal speed and
$\delta B_{mn}$ the perturbed radial magnetic field
with the mode number $(m,n)$. $f_K$ is a factor
introduced to account for the reduction of thermal diffusion due to 
kinetic effects.
%through stochastic field lines due to kinetic effects.
The precise value of $f_K$ is unknown. This will require a
more sophisticated kinetic modelling or a fluid closure 
which is beyond the scope of this paper. In general, we expect
$f_k < 1$, and Ref.~\cite{GYP} suggests $f_K \simeq 0.1$.
In this work, we use $f_K=0.1$ and $1$ for a comparative study 
of the impact of $f_K$ on the parallel energy loss. Main features presented 
in this paper, however, do not change by this $f_K$
variation (except for the amount of parallel energy losses).
We expect the actual value of $f_K$ will be within this range.

Figures~5(a) and (b) show time evolution of instantaneous
heat flux ($Q$) and the cumulative ratio of  
the energy loss to the initial energy ($\Delta W/W$) during simulations. 
Blue (red) lines represents the result when $f_K=0.1$ ($f_K=1.0$).
Also, dotted lines are $Q$ and $\Delta W/W$ due to $E \times B$
convection and solid lines represent those originated from 
$Q_{||}$. When $f_K$ is small (=0.1), the convective loss, $Q_{cv}$, is 
greater than $Q_{||}$ in the early stage of a crash 
({\it i.e.,} when $60 \leq t \leq 80$). 
However, $Q_{||}$ becomes dominant after $t \simeq 80$.
%(i.e. when the island width of PTM is maximized as shown
%in Fig.~3(a)). 
Thus, $Q_{||}$ is responsible for 
the majority of the net collapse-induced energy loss, 
taking up $\sim 75 \%$ of total $\Delta W/W$. 
When $f_K$ is large (=1.0),
$Q_{||}$ is always dominant over $Q_{cv}$, as shown in red lines
in Figs.~5(a) and (b). 
%These results show  hat $f_K=0.1$ is close to the recent 
%estimation showing that the contribution from the filamentary convective
%loss to the total ELM energy loss will be less than $\sim 30 \%$ \cite{Kirk}.
%
%%%%%%%%%
 \begin{figure}
\includegraphics[width=1.0\textwidth]{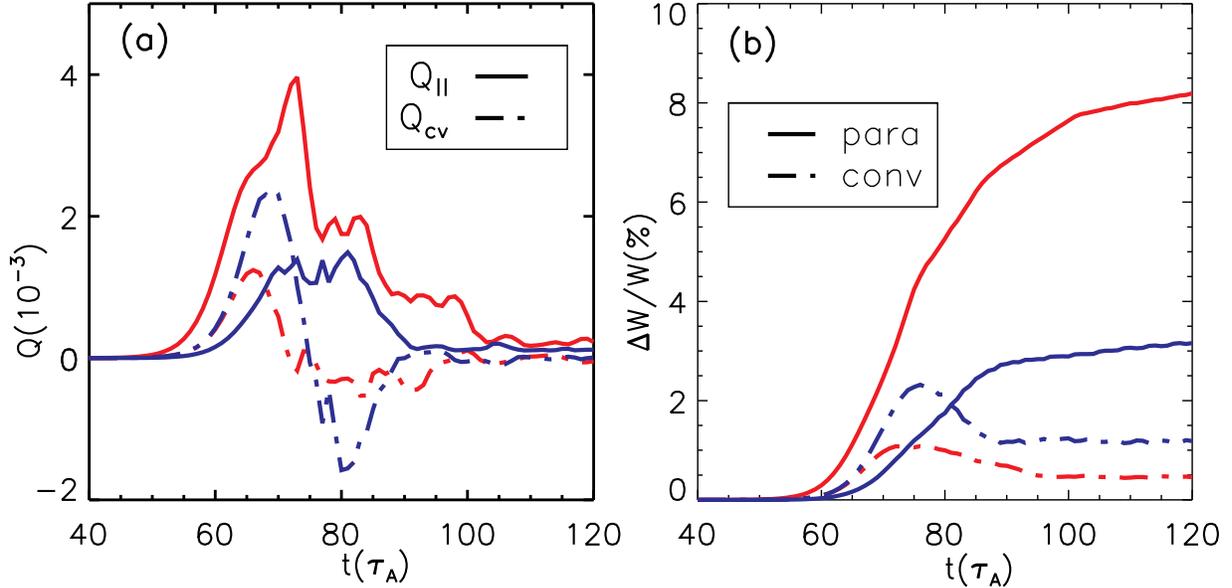} 
\caption{\label{Fig:5} Time evolution of (a) instantaneous
heat flux ($Q$) and (b) the cumulative ratio of ELM energy loss to the initial energy 
$\Delta W/W$ during simulations. 
Blue (red) lines represents when $f_K=0.1$ ($f_K=1.0$).
Dotted lines are $Q$ and $\Delta W/W$ due to the $E \times B$
convection while solid lines represent those 
%originated 
from $Q_{||}$.}
 \end{figure}

%%%%%%%%%%%%%%%%%%%%%%%%%%%%%%%%%%%%%%%%%%%%%%%%%%%%%%%%%%%%%%%%%%%%
%%%%%%%%%%%%%%%%%%%%%%%%%%%%%%%%%%%%%%%%%%%%%%%%%
%% eta_H role %%%%%%%%%%

Now, we briefly discuss the role of hyper-resistivity ($\eta_H$),
which was highlighted in previous studies \cite{Xu-PRL,Xu-NF}, 
in relation to the field line stochastization. 
$\eta_H$ represents electron dynamics in these simulations and may 
originate from, for instance, the residual electron temperature
gradient turbulence in the edge pedestal \cite{Singh}.
%In this study, 
We find that $\eta_H$ plays two key roles in the pedestal
collapse. First, it enhances the growth of STM by increasing the
growth rate of PBM. 
%However, the $\eta_H$ scaling of $\gamma_2^+$ in the simulation,
%$\gamma_2^+ \propto \eta_H^{1/6 \sim 1/8}$, deviates from the prediction
%of an early study reporting $\gamma \propto \eta_H^{1/3}$ \cite{Kaw}. 
%A possible reason for this deviation may be
%due to the fact that STM in the present work is a ``nonlinearly driven mode" 
%growing even when $\Delta^\prime <0$, while the
%tearing mode considered in Ref.~\cite{Kaw} is the conventional one.
Second, it increases the nonlinear
growth of PTM, hence expediting the stochastization process. 
In this way, the increase of $\eta_H$ shortens the ELM crash phase, whilst
the decrease of it delays or even prohibits the pedestal collapse.

In summary, we found, from three-dimensional nonlinear simulations, 
that two important dynamical
processes are involved in the edge pedestal collapse: 
(1) the generation of nonlinearly driven tearing modes from unstable 
ballooning modes and subsequent stochastization of
magnetic fields, (2) the significant parallel
energy loss through the stochastic lines.
%in comparison with convective loss. 
Based on these findings, we propose the ELM crash process as follows:
(1) the generation of STMs through a nonlinear energy transfer between 
adjacent PBMs, (2) the generation of PTMs by extracting energy from a
PBM {\it via} STMs, (3) island overlap, and eventual 
stochastization of magnetic field lines, and (4) a large energy loss
through parallel conduction along the stochastic field lines. 

The present study indicates the possible presence of
a precursor period during which STMs develop from PBMs. 
Strong magnetic activities are then expected
to be observed prior to an ELM crash with a toroidal 
mode number nearly twice as large as that of the original ballooning modes.
%From simulations, 
%We observed, though not shown here, 
%that STMs rotate in electron diamagnetic
%direction, as opposed to ballooning modes which rotate in ion 
%diamagnetic direction. 
In this sense, STMs might be a possible candidate for the 
precursor mode which has been observed in several experiments\cite{KSTAR,Precursor-AUG,Kirk,Precursor-NSTX}.
%To improve simulation model, 
As a future work, it is of importance 
to improve the model for $Q_{||}$ in the presence of stochastic field 
lines by developing a rigorous fluid closure model for it.
Also, studying the impact of external resonant magnetic perturbations
on the pedestal collapse scenario presented in this Letter 
is an immediate next step, which is under consideration.

%%%%%%%%%%%%
%%% ACK
%%%
The authors are grateful to Drs. P. H. Diamond and A. Aydemir
for useful discussions.
They also acknowledge to Drs. X. Q. Xu, A. Dimits, and M. Umansky
for their help in the development of 
analysis codes using the BOUT++ framework.
This work was supported by the WCI program of National Research Foundation of 
Korea funded by Ministry
of Science, ICT and Future Planning of Korea [WCI 2009-001].

\end{document}